\documentclass[]{vgtc}                          

\ifpdf
  \pdfoutput=1\relax                   
  \usepackage{graphicx}                
  \DeclareGraphicsExtensions{.pdf,.png,.jpg,.jpeg} 
\else
  \usepackage{graphicx}                
  \DeclareGraphicsExtensions{.eps}     
\fi%

\graphicspath{{figs/}{Images/}{./}} 

\usepackage{microtype}                 
\PassOptionsToPackage{warn}{textcomp}  
\usepackage{textcomp}                  
\usepackage{mathptmx}                  
\usepackage{times}                     
\usepackage{cite}                      
\usepackage{tabu}                      
\usepackage{booktabs}                  

\usepackage[T1]{fontenc} 
\usepackage[utf8]{inputenc} 
\usepackage{amsmath,cite,url}
\usepackage{color}
\usepackage{enumitem}

\usepackage{caption}
\usepackage{subfig}
\usepackage{capt-of}

\usepackage[bottom]{footmisc}

\onlineid{8694}

\vgtccategory{Application Paper}

\vgtcinsertpkg

\title{CardsVR: A Two-Person VR Experience with Passive Haptic Feedback from a Deck of Playing Cards}

\author{Andrew Huard\thanks{e-mail: ahuard@ucsb.edu}\\ %
        \scriptsize University of California Santa Barbara %
\and Mengyu Chen\thanks{e-mail: mengyuchen@ucsb.edu}\\ %
     \scriptsize University of California Santa Barbara %
\and Misha Sra\thanks{e-mail: sra@cs.ucsb.edu}\\ %
     \scriptsize University of California Santa Barbara}

\teaser{
  \centering
  \includegraphics[width=1\linewidth]{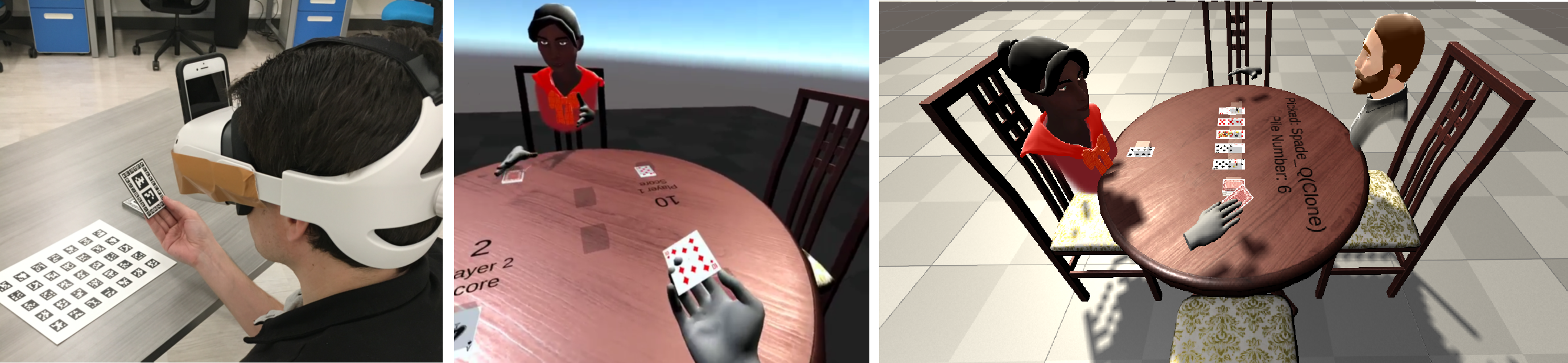}
  \caption{CardsVR is an interactive virtual reality system that enables two participants to play in VR with a physical deck of cards that provide passive haptic feedback. The system allows natural interaction with the physical and virtual objects using hand tracking. Left: a participant picking up a physical card from their deck of tracked cards while wearing a VR headset. Center: the same participant seeing a corresponding virtual card in their virtual hand along with an avatar of another remote player. This second player can be both co-located or remotely located. Right: two-player interaction while playing a virtual card game.}
  \label{fig:teaser}
}

\abstract{Presence in virtual reality (VR) is meaningful for remotely connecting with others and facilitating social interactions despite great distance while providing a sense of ``being there.'' This work presents CardsVR, a two-person VR experience that allows remote participants to play a game of cards together. An entire deck of tracked cards are used to recreate the sense of playing cards in-person. Prior work in VR commonly provides passive haptic feedback either through a single object or through static objects in the environment. CardsVR is novel in providing passive haptic feedback through multiple cards that are individually tracked and represented in the virtual environment. Participants interact with the physical cards by picking them up, holding them, playing them, or moving them on the physical table. Our participant study (N=23) shows that passive haptic feedback provides significant improvement in three standard measures of presence: Possibility to Act, Realism, and Haptics.}

\CCScatlist{
  \CCScatTwelve{Human-centered computing}{Mixed / augmented reality}{}{}
}

\begin{document}

\firstsection{Introduction}
\maketitle

Social connection is fundamental to the human condition.  For the last two years, the COVID-19 pandemic has isolated friends and family from each other, but remote virtual reality (VR) offers great potential to safely connect people while providing a sense of presence or of ``being there," together \cite{Steuer1995}.  Remote presence has been shown to be meaningful for alleviating loneliness, especially in the elderly \cite{Baker2019}, connecting with others \cite{tarr2018synchrony}, and facilitating social interactions \cite{freeman2021hugging} despite great distance, sickness, or disability.  

Touch plays a meaningful role influencing how we experience the world and relate to others \cite{Price2021}. Touch is ever-present in our everyday lives, and a lack of tactile sensation can severely degrade an individual's quality of life \cite{tyson2008sensory}. Neglecting or limiting touch can also diminish the sense of spatial presence in the virtual world (VW) \cite{paterson2006feel}.  Including the sensation of touch in socially interactive VR has been shown to improve the participant's experience and sense of presence \cite{Fermoselle2020}.  While effortless in the real world, replicating the sense of presence in socially interactive VR requires the methodological replication of our bodily senses.  Since the beginning of VR systems in the 1960's, the focus has been on the visual and aural feedback modalities \cite{Sutherland1968}, with touch often being an afterthought \cite{Slater2009}.

Passive Haptic Feedback (PHF) has been shown to improve participant performance on virtual tasks and allow more natural interactions within the virtual environment (VE) \cite{KohliLuv2010}.  Slater \cite{Slater2009} showed that a participant's sense of presence, or of "being there," can break if interactions with the environment feel unnatural.  A disadvantage of current VR interactions using handheld controllers, which rely primarily on the visual and auditory senses, is that they do not recreate the natural experience of picking up or manipulating an object.  In contrast, PHF can improve the sense of presence by engaging the participant's haptic sense and simulating what they can feel and do in the real world with their hands.  
Several examples of PHF have been shown in the literature.  Insko et al. \cite{Insko2001} described passive haptics as "low-fidelity haptic objects" in the physical space that augment a "high-fidelity visual virtual environment" in VR.  A demonstration of PHF is shown in MetaSpace II~\cite{sra2015metaspace2}, which tracks the body and objects in the physical space with corresponding objects in the virtual space.  PHF and tracking of passive objects has demonstrated increasing relevance in commercial systems for instance in manual assembly simulation \cite{Martin2013} and location profile tracking \cite{Chapman2021}.

\begin{figure*}[t]
  \centering
  \includegraphics[width=\linewidth]{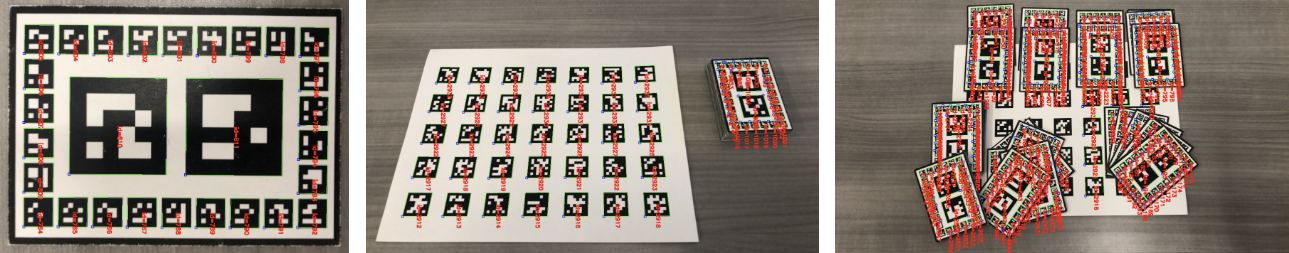}
  \caption{Detected ArUco \cite{Garrido2016,Romero2018} Markers. Left: 28 markers on a single card to track its position and orientation in space. Middle: A board with 35 markers used to calibrate the table at which each player sits to play the VR card game. Right: Multiple cards with multiple self-occlusions detected simultaneously.}
  \label{fig:ArucoDetection}
\end{figure*}

In this work, we present CardsVR, an interactive card playing experience in VR that provides PHF through a deck of physical cards. The VR experience is designed to enable two remotely or co-located participants to play a game of cards as they would in real life.  Co-located participants may wish to enter VR, for example, to experience enhanced gameplay capabilities only available in the virtual space such as manipulating animated objects. CardsVR aims to enhance the participant's sense of presence by introducing PHF using physical playing cards the participant can touch and feel with their hands.  These cards are tracked and mapped to virtual cards, which participants see in VR and interact with the same way they would in real life, with their hands, instead of with controllers.  To the best of our knowledge, ours is the first work that presents PHF with a full deck of cards in an interactive VR card playing application for remotely located participants. The motivation for this work is to create an interactive VR system where participants feel a high degree of presence and that has the potential to benefit non-traditional VR users, such as the elderly, who may be more comfortable using natural interfaces such as their hands rather than controllers.  
We evaluated our system with 23 participants and our results indicate that leveraging PHF to provide a sense of touch in VR can produce a statistically significant improvement in three standard measures of presence as defined by Witmer and Singer \cite{Witmer1998, Witmer2005}: Possibility to Act $(p < 0.012)$, Realism $(p < 0.015)$, and Haptics $(p < 0.027)$.

Our vision is to provide a pipeline for tracking many flat, card-like passive haptic objects that are represented as 3D objects in VR for creating social playful experiences. This also includes non-card-based objects such as board game figures represented physically by flat pieces.
The key contributions of this work are:
\setlist{nolistsep}
\begin{itemize}
    \item{A remote, interactive VR pipeline for PHF through multiple flat objects, each individually tracked and represented in VR.}
    \item{A participant study to evaluate the impact of PHF in a card playing scenario with four themes derived from data analysis on the participant's sense of presence.}
    \item{A PHF system that registers the position and orientation of passive haptic objects one time at the start of gameplay, eliminating the need for continuous tracking of these objects.}
\end{itemize}

\section{Related Work}

\subsection{Bringing Physical Objects into VR}

Passive haptics have been used progressively and at different scales (e.g., room scale\cite{KohliLuv2010}, hand-sized objects\cite{Muender2019}) and materials (e.g., cardboard\cite{Insko2001}, sand\cite{Frohlich2018}).  This work expands on prior systems by exploring the use of PHF through interaction with multiple flat, 2D objects such as playing cards.
Several studies have demonstrated that the sense of presence is maintained even when PHF is implemented through physical objects that approximate what is seen in the virtual world~\cite{Insko2001,sra2016procedurally,sra2017oasis,Muender2019}. In some of these studies, the PHF objects are manually created and matched to an existing VR environment \cite{Insko2001}, while in others the corresponding VR objects are automatically generated using 3D scanning \cite{sra2016procedurally}. 
%
Sra et al. \cite{sra2015metaspace2} explored the design of a roomscale, social experience where multiple participants could manipulate physical objects that were movable and tracked using ArUco markers \cite{Garrido2016}.  Substitutional Reality studied the effect of object fidelity on a participant's suspension of disbelief, ease of use, and level of engagement, and found that large mismatches between physical and virtual passive haptic objects can degrade the participant's sense of realism \cite{Simeone2015}.  Other techniques progressively distort and augment the physical world while anchoring participants within their real-life reality, combining the benefits of both haptic sensing and virtual interaction, and integrating the VW into the real one \cite{Roo2017}.  Our system similarly generates objects in the virtual world that approximately correspond to the physical world with some small variation in represented position due to tracking.  These studies indicate that a participant's sense of presence is not significantly degraded by small variations in shape and position.


These prior works attempt to recreate the physical environment by tracking static objects.  Our system approaches this problem differently.  By knowing the location of physical objects (e.g., the table and cards) and tracking them before gameplay starts, our system does not need to maintain continuous tracking during gameplay.  Prior work shows that perfect one-to-one mapping is unnecessary for achieving a high level of presence in passive haptics systems and our card tracker is inspired by that idea \cite{KohliLuv2010}.  Our work demonstrates that a multitude of objects such as a deck of cards can provide PHF by relying on the assumption that these objects are stationary until the participant interacts with them.

\subsection{Passive Haptics in Multiplayer VR}

A key consideration in VR is the participant's sense of spatial presence or sense of "being there" in the virtual space.  Extended to two-player VR, copresence is the sense of "being there," together with others~\cite{Biocca2002, Durlach2000, Schroeder2008, Gray1977}. Multi-player VR involves sharing a virtual space and participating in joint activities. \textit{Your Place and Mine} allowed geographically separated participants to take a dance lesson together in VR \cite{Sra2018YourPlace}. 
Various family games may be implemented using remote, interactive VR including board games, card games, and puzzles \cite{Minatani2007}. Prior work has demonstrated two-person interaction through large, shared passive haptic objects in the physical environment such as tilting a "wobbly table" ~\cite{Lee2016table}.  Our work, instead, facilitates interactions with small, thin objects that participants can pick up using their hands and share as virtual objects.

\subsection{Tracking Methods in Passive Haptics}

All passive haptics systems require correspondence between the real and virtual worlds for the physical objects the participant will interact with.  One example involves sampling the motion of the VR headset itself for redirected walking, collision avoidance, and interaction with stationary objects in the environment \cite{Azmandian2016}.  VRBox uses a combination of hand tracking and a depth camera to track both the participant's movement and the shape of sand in a real sandbox for controlling a virtual landscape \cite{Frohlich2018}.  Like our system, fiducial markers have been used to track the location of participant-tangible objects in VR\cite{sra2015metaspace2,Lee2004}.  However, unlike other systems, ours does not require the continuous tracking of passive haptic objects.

\begin{figure}[t]
  \centering
  \includegraphics[width=0.85\linewidth]{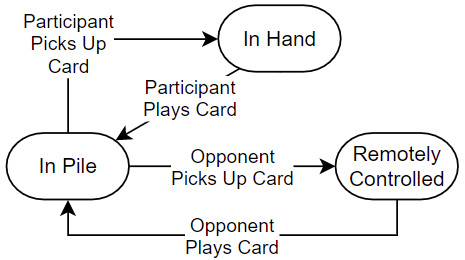}
  \caption{Finite State Machine controlling card object behavior}
  \label{fig:FSM}
\end{figure}

\section{System Design}

Before gameplay begins, players are seated at a table containing several items as shown in Figure \ref{fig:ArucoDetection} (middle):  a deck of custom-printed playing cards that serve as passive haptic objects during gameplay, and a 35-marker board that is used for calibration.  After sitting down and putting the headset on, participants are immediately positioned at a chair on a round virtual table facing their human-player opponent upon launching the CardsVR application.  Both the player and the opponent are represented by avatars (head, body and hands), and both players can see each other seated across the table as shown in Figure \ref{fig:teaser} (right). The player's 1st person perspective during gameplay (center) and their corresponding physical hand movements (left) are compared in Figure \ref{fig:teaser}.  High-fidelity hand tracking, a built-in feature available on the Oculus Quest 2 VR headset, is used to represent the player's hands.  Each player can also see their opponent's hands, which are similarly tracked using the inside-out tracking on the Oculus Quest 2 device.  When players put their hands on the physical table they can see their hands resting on top of the virtual table, which has 1-to-1 correspondence due to calibration.  Players can communicate with each other using voice, which is recorded, transmitted over the internet, and played through the speakers on the remote player's headset with a voice communication latency of approximately 0.5 seconds.  The VR headset provides the ability to specify a visual boundary or guardian that alerts the wearer as they approach obstacles such as walls in the room.  Although a stationary guardian was used during the study to calibrate the headset's floor level, the game is compatible with a roomscale guardian and allows participants to move to their opponent's side of the table, though it was not used during evaluation.

The VR experience offers two modes of play: either with PHF provided by physical cards corresponding to virtual cards in the environment or only virtual cards using 
the game's finite state machine mechanics to enable gameplay.  Virtual cards are objects the participant can see and interact with in the virtual world, which correspond to a tangible card object in the physical world that the participant can manipulate with their hands.  The opponent's cards are virtual objects without a corresponding object in the local player's real world space. The opponent and their cards are remotely located.  It is possible for the local participant to pick up remotely placed cards if a physical card is available in that location on the tabletop.  For example, the opponent may place a remote, virtual card in the local participant's pile, which may already contain a physical card in the real world that the local participant had placed earlier in their turn and that would allow them to pick it up at their next turn. 
Participants interact with physical and virtual objects with only their hands.  The only interfaces in the system are an auxiliary iPhone camera for tracking card location and hand tracking provided by the Oculus Quest 2 headset.  Oculus Quest 2 controllers are not used.  The iPhone is affixed to the front of the VR headset.

\begin{figure}[t]
  \centering
  \includegraphics[width=0.75\linewidth]{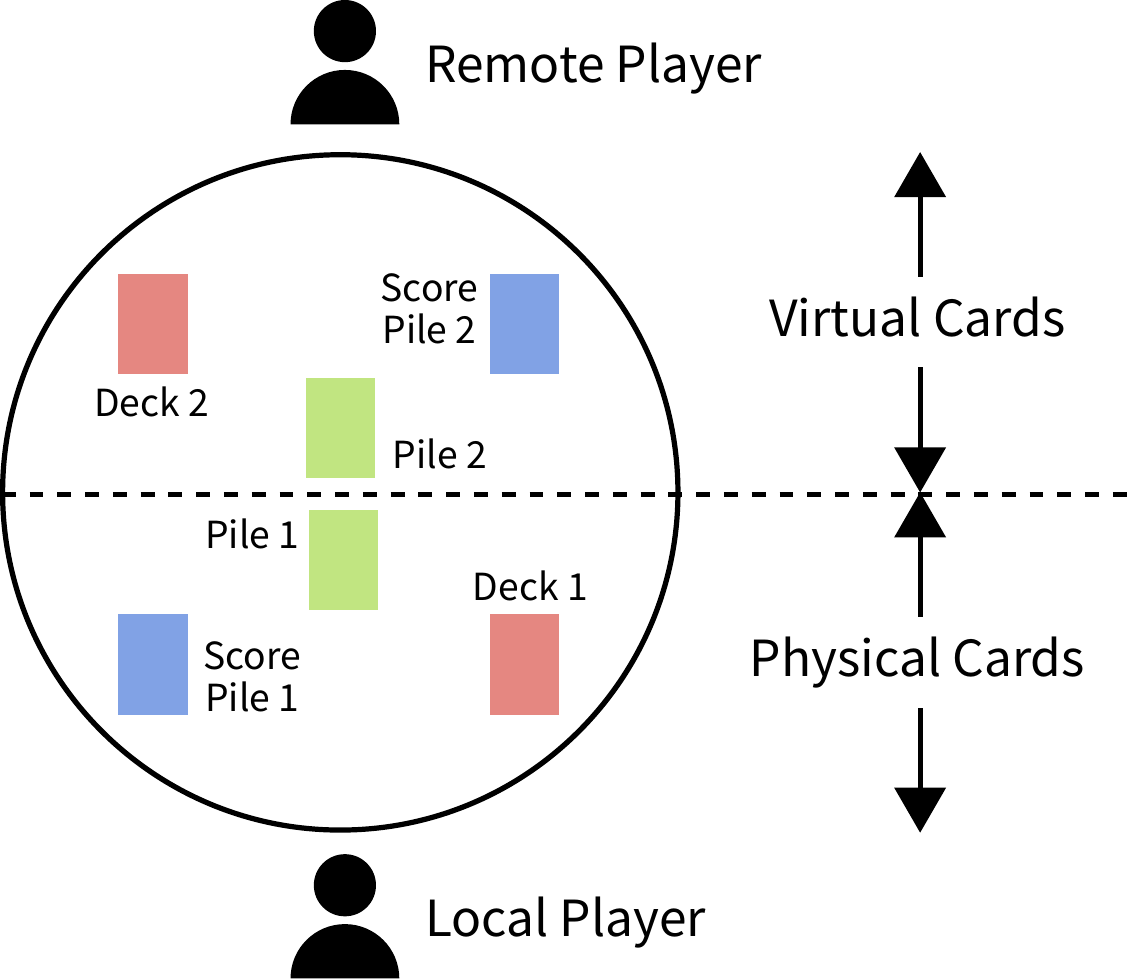}
  \caption{Card locations as seen from the perspective of the local player when using physical cards.  Each participant interacts with physical cards (Decks 1 and 2) on their side of the table and a mix of physical and virtual cards in the middle (Piles 1 and 2).}
  \label{fig:PileDefinition}
\end{figure}

\subsection{Implementation}

The CardsVR system was built using the \textit{Unity} game engine and was implemented using a pair of \textit{Oculus Quest 2} devices and two \textit{Apple iPhones}.  Custom software was built and compiled for \textit{iOS} and installed on iPhone devices which were used to track the cards for providing PHF.  Each iPhone was attached to the front of each player's \textit{Quest} headset to provide an ego-centric viewpoint of the card table as seen from the rear iPhone camera.  Multiplayer data and voice connectivity was implemented using \textit{Photon Engine's PUN, Realtime, and Voice} frameworks\footnote{\url{https://www.photonengine.com/}}.

Upon starting the VR application while seated at a physical table, players are presented with a deck each of virtual playing cards, which correspond to a deck of 52 physical cards on their table.  When a virtual card is picked up, it snaps into the participant's hand, following the hand until the card is placed down on to pre-defined table (pile) locations denoted by transparent blocks. These pile locations correspond to a players personal deck of cards placed closer to them and two playable stacks near the middle of the table as illustrated in Figure \ref{fig:PileDefinition}.  The deck is the only pile location that is pre-populated at the start of gameplay; the other piles start empty.  Players are free to move their cards, one at a time, from any pile to another including to and from the initial deck.  The player's score and the opponent's score are shown next to their score pile locations. We implemented the two-person card game of War, which is a game of random chance. It starts with one player drawing a card from their deck and placing it face up in the middle of the table for the other player to see. The second player draws a card from their deck and places it next to the first card.  The player with the higher ranked card wins the turn and captures both cards in the middle piles and places them in their score pile, following which the next turn starts.

\begin{figure*}[t]
  \centering
  \includegraphics[width=0.9\linewidth]{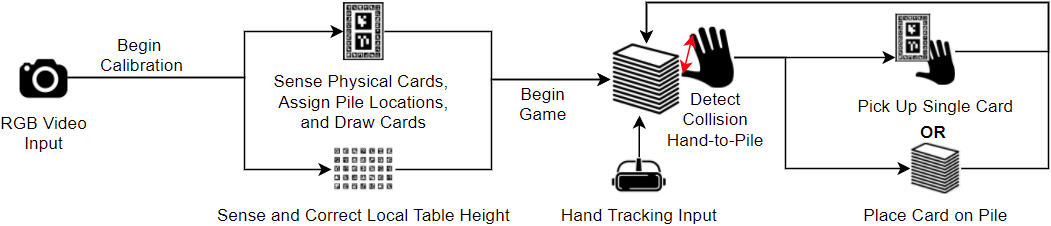}
  \caption{Overview of the CardsVR System (from left to right). Pile locations and the table height are calibrated before gameplay begins.  The participant picks up or places down physical cards using their hands as inputs, which moves corresponding virtual objects.}
  \label{FigurePipelineSystem}
\end{figure*}

The game uses a finite state machine (Figure \ref{fig:FSM}), to control whether each card is classified as being held in the participant's tracked hand, is in a pile, or is being manipulated by the remote player.  Cards transition state whenever the hand passes through an object collider, which notifies observers attached to the cards to change state.  After a card changes state, another state change is prevented until at least 2 seconds have passed, which minimizes rapid toggling of hand-pile states from occurring in a single participant action.  An overview of the system is presented in Figure \ref{FigurePipelineSystem}.  User evaluation revealed that this two second blocking interval should be extended to a longer period to allow for more time to pick up physical objects. 

\subsubsection{Passive Object Tracking}\label{sec:passiveObjTrack}

Our system tracks the position and orientation of physical cards using ArUco tags \cite{Romero2018, Garrido2016} printed on both sides of a full deck of blank playing cards.  Each card has 28 unique markers per side (26 small and 2 large), as shown in Figure \ref{fig:ArucoDetection} (left), for a total of 2,912 unique ArUco tags per deck.  A 35-marker letter size board is used to track and calibrate table height.  Three ArUco dictionaries were used to source up to 3,000 tags at a fixed rate of 30 FPS.  The tag IDs, corners, and screen dimensions were transmitted to the headset for further processing and rendering. 

Marker locations on each card were selected for improving detection when occluded and to reduce tracking errors with redundancy.  
In our setup, only one marker out of 28 needs to be unobstructed to successfully track the location of each card.  An extreme example showing tracked markers in the presence of card occlusion is illustrated in Figure \ref{fig:ArucoDetection} (right).  Our application begins with the starting arrangement of Figure \ref{fig:ArucoDetection} (middle), which consists of a single stacked deck and an unobstructed table calibration board.  Once gameplay starts, the tracker is turned off and the cards can be strewn around the table as shown in the Figure \ref{fig:ArucoDetection} (right) with no impact on performance.
On the headset, the tag IDs are looked up in a database, which provides a fixed offset vector from the tag center to the card center based on the tag's fixed position on the card.  These offset vectors are calculated offline when the card designs are first created and stored for later reference by the tracking algorithm.  Each tag produces a corresponding card model candidate using this offset vector. The offset is applied to the card model before transforming its position and orientation into world coordinates. 

When the game launches, the card and table trackers are enabled and the virtual card deck location and table height are corrected to match the location of their corresponding physical objects.  The table tracker is disabled by turning over or removing the 35-marker board from the camera's field of view.  When the player picks up a physical card from the deck for the first time, the card tracker is turned off to avoid position jitter, which is discussed in Section~\ref{sec:Calibration}.  Physical cards are connected to the virtual cards through 1-to-1 position and orientation correspondence in 3D world coordinates, which is maintained throughout the game.  When the player moves a physical card, they see the corresponding virtual card moving to the same position.  If they want to pick the card up again later  they can see where they left it in VR.  

\subsection{Continuous Tracking and Jitter}
Our results, described in Section \ref{sec:Results}, shows that continuous tracking is unnecessary for improving the sense of presence with passive haptic objects.  Our system demonstrates that 1-to-1 position correspondence can be maintained without continuous tracking, which would otherwise introduce jitter that degrades the user experience.  Instead, the movement of these passive haptic objects is determined indirectly through other control mechanisms such as tracking the participant's hands.  We empirically determined the experience of passive haptic interaction using physical cards with and without continuous tracking.  As reported in prior work \cite{Batmaz2019}, continuous tracking resulted in positional and rotational jitter that interfered with immersion and sense of presence.  

It has been shown that rotational jitter in excess of $\pm0.5^{\circ}$ degrades participant's throughput in manipulating objects and increases error rates~\cite{Batmaz2019}.  One commonly used approach to reduce apparent jitter is to use low-latency, extended Kalman filters to accurately predict an object's position and rotation to reduce noise (jitter) \cite{Welch2009}.  
Reducing tracker jitter (noise) is challenging in passive haptics systems that track many objects simultaneously due to the need for low-latency filtering, which requires additional processing time and limits the system's attainable Frames Per Second (FPS). 
Without filtering, our approach is capable of tracking position and rotation of all 52 cards in a deck, with low latency (greater than 30 FPS) while providing a comfortable and immersive experience.

\subsubsection{Calibration}\label{sec:Calibration}
Calibration of the table height and card deck positions are performed prior to the start of gameplay (Figure \ref{fig:ArucoDetection}, middle). 
Once play starts, the tracker is turned off to avoid position and orientation jitter, which can cause discomfort and reduce the user's sense of presence.  Disabling the tracker prevents the correction of minor positional correspondence errors such as cards sliding on the table or being pushed slightly once placed on a pile.  We found that these correspondence errors (2cm or less) did not interfere with gameplay, and participants were able to pick up physical cards once they were placed down.  Prior work has shown that passive haptics do not require strict 1-to-1 mapping between real and virtual objects \cite{KohliLuv2010, Simeone2015}.

Table height is calibrated in VR using a 35-tag grid of ArUco markers, which are placed on the physical table surface.  The card deck positions are calibrated using ArUco markers printed directly on the cards, which are then projected onto the table's surface plane in VR. The calibrated deck locations are synchronized so that the player sees the correct correspondence between the opponent's hands and cards being drawn from their deck.  Cards are then dealt to these locations by stacking the virtual cards in VR to locations corresponding with physical cards the player can touch and manipulate.  The system maintains 1-to-1 correspondence between the real and virtual worlds after calibration without the need to continuously track card and table objects during gameplay. 

Using a 35-tag marker grid calibration board, the iPhone camera's device dependent translation and rotation offset from the headset origin is calibrated offline by manually adjusting the offsets until the 3D reconstructed output matches the corresponding position and origination of the calibration board as seen by the operator wearing the VR headset.

\subsection{Tracking Algorithm}

\begin{figure*}[t]
  \centering
  \includegraphics[width=0.9\linewidth]{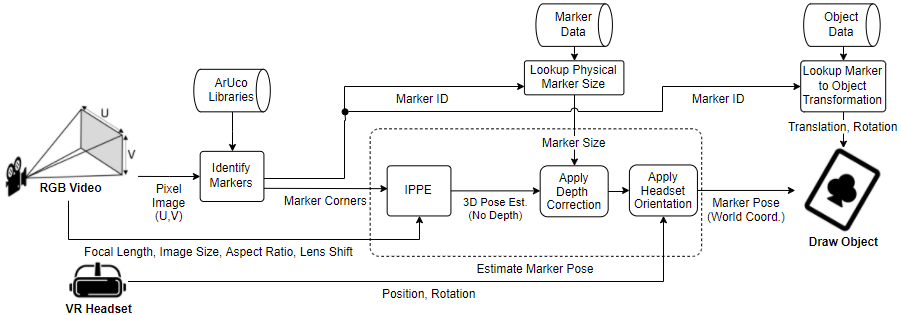}
  \caption{The position and orientation of tracked objects is represented in the virtual space by estimating the pose and depth of fiducial markers attached to that physical object and then applying a translation offset.}
  \label{FigurePipelineTracker}
\end{figure*}

The tracker pipeline is illustrated in Figure \ref{FigurePipelineTracker}.  Depth is controlled using a constant scaling factor term (s), which is first adjusted to achieve 1-to-1 physical-virtual correspondence for the smaller tags positioned around the periphery of the cards.  This scaling factor is then adjusted for the larger tags by multiplying the scaling factor (s) by the side-length ratio of the different size tags.  Since the smaller tags have a length of 0.3 inches on edge and the larger tags have a length of 0.9 inches on edge, the scaling factor for the larger tags is $0.9s/0.3=3s$.  Therefore, the tracker's depth engine is controlled by a single scaling factor that is adjusted for tags of different sizes.  This depth scaling factor is used to control the size of the tag's canonical model.  Details of the tracking algorithm are included in Section S1 of the supplementary files. 

\section{Evaluation}

\subsection{Participants}

We invited 32 participants in pairs (10 female, 22 male) to test the CardsVR system, which included 8 participants in a pilot group (5 female, 3 male).  Data from the pilot group 
were excluded from the statistical analyses described below. The other 24 participants were recruited from a university campus in North America and members of the broader community.  One of these participants did not complete the required questionnaires and their data was eliminated from analysis.  Therefore, questionnaire data from 23 participants out of the 32 total were eligible for statistical analysis and evaluation.  

\paragraph{Pilot Study}

The pilot study was conducted off-campus to test the system and to develop and refine participant study questionnaires.  Pilot group feedback provided from the free-form questionnaires were retained for reporting in the inductive thematic analysis (N=32).  However, 1-7 point Likert scale presence survey data obtained from the pilot group was excluded from the statistical analysis. 
The results of the pilot group's free-form questionnaire were retained as the feedback was specific to the descriptive aspects of the VR experience.

\paragraph{Groups 1 \& 2}

The eligible participants were separated into two groups.  Each group was either provided a deck of physical playing cards (Group 1, N=12, "With PHF") or played the game with only virtual cards (Group 2, N=11, "Without PHF"). Of the 23 eligible participants, 7 were 18-24, 9 were 25-29, 4 were 30-39, 2 were 60-69, and 1 was 70-79 years of age.  The mean age of eligible participants was 32 years.  
Groups 1 and 2 performed the study in an on-campus research lab. 

\subsection{Apparatus and Task}\label{section:GameSelection}
We used two separate Oculus Quest 2 devices and two Apple iPhones, one for each participant for the study. The two participants were seated at tables away from each other in two separate rooms with a soundproof wall separating them such that they could not see each other to simulate remote participation. The participants could not directly communicate and instead spoke with one another through the VR game. Participants experienced playing the two-person game of War. 
%
%
The game of War is a game of random chance where players draw a card from their deck and place it in the middle of the table facing their opponent. The player with the higher ranked card wins the battle,  captures both cards, and places them in their score pile to be counted as points. The game continues until all cards in the deck have been played. This game was trivially simple to play and easy to teach, which mitigated any potential risk of game difficulty biasing the participant study.

\subsection{Procedure}

Our study aimed to investigate the effect of PHF on the participant's sense of presence in the VR experience.  Following calibration and before starting, participants were instructed on the rules of the implemented card game and in the use of physical cards during VR gameplay.  To compare the experience with and without using PHF, we used a between-subjects design. Participants played the card game either with (Group 1) or without (Group 2) the physical cards. All participants used hands instead of VR controllers to interact with the virtual cards.  The \textit{with PHF} group could physically pick up cards from their own deck, and could see their opponent's remotely located cards when they were played in the middle of the table.  Players could also pick up the opponent's cards with or without a corresponding physical card depending on turn-taking in the game. 

\subsubsection{Questionnaire}

After each study session, participants were asked to complete a questionnaire to collect feedback on their experience, which consisted of several parts:\\

\noindent\textit{Presence Survey}\\
A tailored presence survey rated on a 1-7 point Likert scale.

    \begin{itemize}
        \item Questions were taken from Witmer and Singer's~\cite{Witmer1998} presence questionnaire, which included 19 core questions from version 3.0 along with an additional three auditory and two haptic items from version 2.0.
        \item For analysis, we classified Witmer and Singer's~\cite{Witmer1998} 19 core and 5 additional questions into 7 categories as was done by Robillard et. al.  \cite{robillard2002validation}.  We excluded both (2) questions from the auditory category because the category was not applicable to the study, which left 6 categories as listed in Section S2 and Table S1 in the Supplemental Material.
        \item Four out of the remaining 22 source questions were removed as they were unrelated to the experience being evaluated (See Supplemental Table S1).\\
    \end{itemize}

\noindent\textit{Custom Questionnaire.}
A second custom questionnaire was taken consisting of 10 open-ended questions (see Supplemental Table S2 and Section S3 for a detailed listing of questions including the prompts).

\subsubsection{Ethical Approval}

This participant study was granted prior ethics approval for the use of human subjects by the UCSB Office of Research.  Participation was voluntary, and participants were free to refuse or withdraw from the study at any time. An information sheet was provided prior to obtaining informed consent to participate. Participants were offered \$15 as compensation for their time.

\subsection{Results}\label{sec:Results}

Of the 23 scored participants, 12 played with PHF (Group 1) and 11 played without PHF (Group 2). 
To evaluate the effect of PHF on a participant's sense of presence, a series of significance tests were performed.   

\subsubsection{Preliminary Tests for Equal Variance and Normality}

Bartlett's Test \cite{Bartlett1937} was performed to determine the homogeneity of variance between sample sets.  Between Groups 1 and 2, comparing survey responses with and without PHF, the results of Bartlett's test revealed homogeneous variance among three categories: Possibility to Act $(p<0.002)$, Realism $(p<0.018)$, and Haptic $(p<0.049)$.  After excluding research engineers from the same sample sets for reasons described in section \ref{section:Discussion}, Bartlett's test failed to establish homogeneous variance in the Haptics category $(p<0.344)$. Shapiro-Wilk tests were then performed to evaluate the sample sets for normality.  The null hypothesis was rejected between groups (labeled A and B in Table \ref{tableMannWhitney_2}) in all six categories $(p>0.05)$.  Therefore, Gaussian-Normal distributions cannot be established.  These results constrained the statistical significance tests available, which we describe in section \ref{section:MannWhitneyTest}.  The number of samples in each category and group (A, B) differ because the categories contained varying numbers of questions and some participants did not answer every question in the survey.

\subsubsection{Identifying Interactions and Factors of Significance}\label{section:MannWhitneyTest}

For normally distributed sample sets when homogeneity of variance cannot be established, Welsh's ANOVA may be performed, which tests the probability (p-value) of both samples sharing the same source population \cite{liu_2015}.  Wilcoxon-Mann-Whitney (WMW) testing \cite{MannWhitney1947} may be used in place of Welsh's t-testing where Gaussian normal sample distribution cannot be established \cite{FayWilcoxonMannWhitney2010}.  
A set of WMW analyses were performed to determine factors of significance and to identify interactions between factors affecting a participant's sense of presence.  These factors include prior VR experience level, passive haptics use (Groups 1 or 2), and education level.

\section{Inductive Thematic Analysis}
A free-form questionnaire, with prompts listed in Table S2 in the Supplemental Material, was provided to participants separately after they rated their sense of presence (Witmer \& Singer\cite{Witmer1998,Witmer2005} on a 1-7 Likert scale (1=Strongly Disagree, 7=Strongly Agree). We asked participants, N=32 including the pilot group, for feedback on the usefulness of the system, its weaknesses and their overall experience using it to play cards in VR with others. We conducted an inductive thematic analysis \cite{braunclarke2006} on the questionnaire data. This consisted of tallying the participants' answers into themes to reveal the most frequently reported feedback. Two researchers independently reviewed the questionnaire answers and derived their own codes and descriptions of the codes. A discussion between the two researchers was held to refine their codes and jointly arrive at the most common themes and categories. The top four themes were gameplay experience, virtual and physical interplay, natural and easy hand interaction, and sense of presence with others.  Participant identification numbers and question numbers (Supplemental Table S2) are referenced herein using a shorthand notation.  For example, participant \#15 responding to question \#4 is referenced as "P15/Q4."

\subsection{Theme 1: Gameplay Experience}
The following section describes how participants experienced CardsVR gameplay.  Participants' responses were divided into four prominent categories: Sense of Enjoyment, Anticipation, Ease, and Focus and Self-Control.
\\
\\
\noindent\textit{Sense of Enjoyment.}
Many participants (15/32) expressed that the experience was enjoyable, fresh, and novel in bringing a classical game into the virtual space.  Participant \#25 wrote in response to Question \#3 (using the shorthand notation P25/Q3), "Playing cards in VR was a pretty fun experience.  It was pretty straightforward, but the VR itself was novel."  One participant changed their initial impression of the concept after the experience of playing cards with our system (P14/Q10), "Skepticism of the concept. Not easy to appreciate why this is as fun as real cards until you do it."  Several participants commented on how fresh the experience felt, noting (P8/Q3) that it was a "different kind of experience. I would do more," and (P17/Q4), it was a "unique experience with a familiar game."
\\
\\
\noindent\textit{Sense of Anticipation.}
Several participants (4/32) noted a desire to play a greater variety of games such as Texas Hold'em (P17/Q3) and Poker (P4/Q8).  Participants noted how virtual card games could be extended into a larger variety of games, writing (P21/Q8), "Many great games could be modeled by this like Magic the Gathering or Monopoly." Another participant had an even more abstract vision for future applications of the system, writing (P17/Q2), "Can program new card games that would be difficult or impossible to reproduce physically."
\\
\\
\noindent\textit{Sense of Ease.}
Participants showed little trouble learning how to use the system for the first time.  One participant wrote (P28/Q5), "I don't think there was a hard part as such. The instructions were easy to understand and it was a fun game." Another commented (P2/Q8), "it's a universal type of game that anyone can quickly pick up and play."
\\
\\
\noindent\textit{Focus and Self-Control.}
Participants reported that playing card games in VR enhanced their sense of focus and self-control.  One participant noted an improvement in focus and concentration (P22/Q9), "It's a more immersive experience where less distractions are present."  Another participant wrote (P11/Q7), "It was nicely laid out and having the cards easily organized alleviated stress."

\subsection{Theme 2: Virtual and Physical Interplay}

Participants praised the ability to bridge their physical reality with a virtual environment, including cards, tables, and their hands.  Some participants (5/11) indicated that the correlation between the two provided them a good sense of presence and realism including an improved sense of embodiment in the virtual space.  Participants' responses were divided into three  categories: Melding of Physical and Virtual Environments, Game Mechanics and Expectation for Continuous Tracking.

\begin{table*}[t]
\centering
\begin{minipage}[b]{.70\textwidth}
 \resizebox{1\columnwidth}{!}{%
      \begin{tabular}{cccccccc}
        \toprule
         & Possibility &  & Possibility & Self-Evaluation & Quality of & Passive \\
        Statistic & to Act & Realism & to Examine & of Performance & Interface & Haptics  \\
        
        \midrule
        P-Value & \textbf{0.012*} & \textbf{0.015*} & 0.505 & 0.824 & 0.121 & \textbf{0.027*} \\
        Average (A) & 5.66 & 5.23 & 5.06 & 6.00 & 5.92 & 5.00 \\
        Average (B) & 4.91 & 4.52 & 5.47 & 5.88 & 5.07 & 4.06 \\
        Std Dev (A) & 0.814 & 1.06 & 1.48 & 1.12 & 1.00 & 0.87 \\
        Std Dev (B) & 1.27 & 1.37 & 1.19 & 1.05 & 1.74 & 1.11 \\
        N Samples (A) & 32 & 40 & 16 & 8 & 24 & 16 \\
        N Samples (B) & 35 & 44 & 17 & 8 & 27 & 17 \\
        Effect & 0.711 & 0.578 & 0.305 & 0.115 & 0.616 & 0.952 \\
        Power & 0.817 & 0.742 & 0.136 & 0.055 & 0.576 & 0.755 \\
        \bottomrule
        \end{tabular}
        }
        \caption{Wilcoxon-Mann-Whitney Test with PHF (A) and Without PHF (B), excluding the subgroup of 6 research engineers.  *Note: Significant values $(p<0.05)$ are noted in bold print.}
        \label{tableMannWhitney_2}

\end{minipage}\qquad
\begin{minipage}[b]{.25\textwidth}
  \includegraphics[width=1\linewidth]{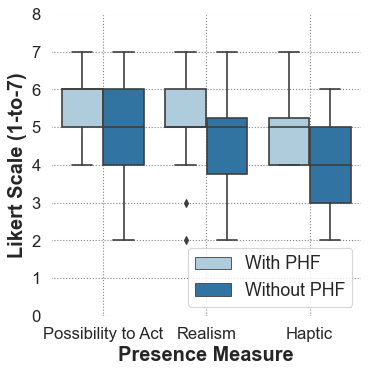}
  \captionof{figure}{Ratings excluding the subgroup of 6 research engineers.
  }
  \label{BoxPlot1}
\end{minipage}
\end{table*}

\noindent\textit{\\Melding of Physical and Virtual Environments.}
5/11 participants in Group 2 reported a sense of physical immersion due to the system's use of hand tracking and anchoring by surrounding static objects in the virtual environment despite not interacting with physical cards.  For example, one Group 2 participant (P28/Q4) noted that the most appealing aspect of the experience was being "able to play without having to use the controls in each hand. That made me feel more free, since I wasn't afraid of dropping them. I also liked the fact that I was sitting in a chair with a table in front of me... it made me feel I was really inside the game."  In contrast, 5/12 participants in Group 1 mentioned their interactions with the physical cards as being the most appealing aspect of the experience.  For example, participants wrote that the most appealing aspect for them was (P2/Q4), "not having to shuffle the cards," and (P11/Q4), "easy cleanup and no need for shuffling."
\\
\\
\noindent\textit{Game Mechanics.}
12/32 participants mentioned that some gameplay mechanics were different from the real world experience of playing cards.  Participants noted that the aspect they liked the least was, "[t]he fact that you could only move one card at a time, and the slow pace of play" (P16/Q6). Participants also felt the fixed locations of the card piles in the virtual space limited their sense of realism.  According to one participant (P22/Q1), "The collision boxes felt too game-like and not real enough."  Although participants were instructed how to play the game and interact before the experiment began, they noted some initial difficulty figuring out the game mechanics after entering VR, with one participant commenting (P15/Q6), "I would’ve liked some more visual aids."  This indicates that better indicators of card piles than the visible collision boxes would be more helpful.
\\
\\
\noindent\textit{Expectation for Continuous Tracking.}
6/32 participants stated their expectation for continuously tracked cards, which the system does not provide.  Cards are tracked during the initial setup and calibration of the scene, then the card tracker is turned off during gameplay to avoid position and rotation jitter that could otherwise degrade immersion.  The system instead relies on game mechanics and the headset's built-in hand tracking to move the cards by attaching them to the player's hand.  Although the system has a timer that delays attachment of the card to the hand while the user picks it up, some participants may pick up the physical card too slowly, resulting in the game mechanics placing the card in the user's hand before it is fully picked up.  According to one participant (P20/Q5), "To remind myself to pick up the physical card when I visual see the card already sticking on my fingers. My mind is tricked."  This feedback indicates that change in physical-to-virtual synchronization can degrade a user's sense of immersion and even trick their mind to do actions that are unintended.

\subsection{Theme 3: Natural and Easy Hand Interaction}

10/23 participants indicated that hand-only interaction was better than using a traditional VR controller because it is more natural and intuitive.  Hands-only interaction provides an easy interface that the participant knows how to use right away instead of having to learn buttons on a controller.  Participants' responses were divided into two prominent categories: Using Hands as Controllers and PHF.
\\
\\
\noindent\textit{Using Hands as Controllers.}
10/32 participants were eager to use their hands as controllers noting how natural it was instead of using input devices.  One stated (P20/Q9), "It is more natural and intuitive to interact with my hands than using a mouse."  Participants also were surprised that they could interact with the physical and virtual environments simultaneously using their hands.  Another commented (P4/Q7), "I was surprised that I could control cards with only my hands."  These responses indicate that using hand tracking in VR applications for interacting with physical objects can be intuitive, natural, and easy to learn.
\\
\\
\noindent\textit{Passive Haptic Feedback.}
Group 1 participants (8/12) also welcomed the ability to play in VR with real physical card objects they can hold in their hands, which added to the sense of presence.  For example, a Group 1 participant  described (P26/Q4), "The most appealing experience was a physical card being correlated to what you see in virtual reality, allowing for the card to be manipulated with the hand."  Another Group 1 participant (P8/Q7) said they were, "surprised how easy it was to pick up cards."
In contrast, participants from Group 2 (5/11) seemed to lack the same sense of control as Group 1.  For example, a Group 2 participant wrote (P4/Q5), "I wasn't able to control the hands all the time. Reaching was a problem."  Another participant commented (P13/Q5), "Getting used to the movements required."  When asked what was the hardest part of the experience, one participant replied (P12/Q5), "The distance and touch commands." Group 2 participants had to rely solely on game mechanics to move the cards virtually, without the benefit of having physical cards they could hold in their hands.  This indicates that PHF made the gameplay experience more controllable.

\subsection{Theme 4: Sense of Presence With Others}

15/32 participants reported that the virtual environment felt more real to them because of their remote interactions with a virtual partner avatar. Interactions such as picking up a card from a partner in VR can add an extra sense of presence and immersion.  Participants also expressed that they see playing card games in VR helping them connect and socialize with friends and family.  Responses were divided into three categories: Presence Enhanced by Seeing Other's Actions, Remote Interaction, and Social Interaction.
\\
\\
\noindent\textit{Presence Enhanced by Seeing Other's Actions.}
According to 15/32 participants, seeing the motion of other players can provoke a social-emotional response due to increased sense of presence with others.  For example, a participant responded (P3/Q1), "Seeing the motion of the cards and the other player provided a sense of their emotions (enthusiasm, hesitation, excitement)."  The avatar can be rendered in low fidelity and still maintain a high degree of social presence.  For instance, one participant noted (P29/Q1), "I felt like I was interacting with an other human, even if the avatar wasn't the most realistic. The dynamics of the game, plus the possibility to chat to the other participant made me feel comfortable in my VR skin."
\\
\\
\noindent\textit{Remote Interaction.}
Most participants (20/32) expressed a desire to connect remotely with others.  For example, a participant (P25/Q2) described those who could benefit from remote card games interaction as, "People who live far from their family or friends, or maybe people who have to quarantine but want some interaction."  Several participants also noted a desire to extend the card game to other forms of remote activities such as board games, meetings in VR, or even developing documents.  For example, participant (P21/Q2) described, "Imagine in the future whole board games can be played or plans be discussed over VR, where all can edit a blueprint in real life."
\\
\\
\noindent\textit{Social Interaction.}
A few participants (3/32) noted mixed feelings about the social aspects of VR card games.  One participant cautioned (P16/Q8), "I think cards is at its most fun in a raucous, social environment. VR seemed more subdued and isolating."  Another participant noted (P15/Q10), "I think people are still uneasy about it replacing real interactions."  Other participants welcomed social interactions in VR, for example one participant (P28/Q8) wrote, "It really made me feel like I was casually playing with someone else that was there in VR but not really in front of me.  I see myself playing VR card games with my friends in the future."  These mixed results indicate that the social engagement expectations varied among the participants, 
and were likely based on their prior card playing social experiences and perhaps the degree of COVID-19 induced degree of social isolation.

\subsection{Discussion}\label{section:Discussion}

PHF using physical cards was shown to improve the participants' sense of presence in three areas:  Possibility to Act, Realism, and Haptics.  
This indicates that participants feel like they are actually holding cards while interacting with passive haptic objects in-game.  Qualitative feedback confirms these results where participants reported feeling like they were ``playing cards in real life'' while experiencing PHF.  
Ratings in the Possibility to Act measure also showed a strong improvement (+0.75 mean increase).  This presence measure assesses the participant's sense of ability to perform actions within the environment, which was enhanced by using physical cards.  The participant can do several real world actions while holding the physical cards that they cannot do without them such as picking them up, dropping them, and feeling them in their hands.  Improved sense of presence has been shown using hand tracking versus hand-held controllers \cite{Moon2022}.  However, all participants in our study used controller-free hand tracking only.  Our results indicate that PHF provided further enhancement in the sense of presence over hand tracking alone.    

We noted that the subgroup of research engineers (N=6) rated passive haptic use differently than the other non-engineer participants (N=17). We theorize that this subgroup was focused on the technical aspects of the design such as examining perceived imperfections of the tracker or lack of continuous tracking that largely went unnoticed by the other participants.  Figure \ref{BoxPlot1} shows the distribution of scores with and without PHF when research engineers are excluded (Table \ref{tableMannWhitney_2}).
After excluding them from the samples, the results show that PHF provides a significant improvement in average subjective ratings in the areas of Haptics (+0.94 mean increase, $p<0.027$), Possibility to Act (+0.75 mean increase, $p<0.012$), and Realism (+0.71 mean increase, $p<0.015$).  The results of WMW testing including all subgroups is shown in Figure S1 and Table S3 in the supplemental material.

It is possible the research engineers were distracted by focusing on the technical aspects of the experience, which interfered with their attention to and sense of presence in the VR experience.  Their average subjective ratings indicate degraded results with PHF in the areas of Haptics (-0.62 mean decrease, $p<0.427$), Possibility to Act (-0.88 mean decrease, $p<0.036$), and Realism (-0.35 mean decrease, $p<0.282$).  They expressed more technical curiosity about the tracking components as seen in the open-ended feedback. For example, P30 wrote, "Tracking was limited. The execution of tracking needs to be improved. With improvements in tracking, I think it would provide an experience very close to the real physical interaction." Interestingly, the reduced sense of presence due to lack of continuous tracking was the opposite of what has been shown in prior work \cite{Batmaz2019} where jitter caused by continuous tracking was the reason for a reduced sense of presence.

We rule out an alternate hypothesis that the results were influenced by less novelty in hand tracking among engineers having advanced familiarity with VR.  A majority of both engineers (50\%) and non-engineers (60\%) reported being "somewhat familiar" with VR.  Significant distributions ($p<0.05$) could not be found when excluding those with moderate to advanced familiarity.  Therefore, hand tracking familiarity does not appear to be the root cause for research engineers rating our system differently than their peers. 

\subsection{Limitations and Future Work}

Our system does not integrate continuous tracking of PHF objects given the challenges in successful implementation for such a large number of objects as used in our system without degrading the user's experience. While our solution is significantly simpler and computationally less expensive with much lower latency, as successfully demonstrated in use by other VR applications such as Auto Hand\footnote{\url{https://www.earnestrobot.net/unityassets/autohand}}, without continuous tracking the cards are not positioned in the participant's hands with 100\% match between the physical and the virtual cards, which one participant (P23/Q3) reported, ``threw off'' their sense of immersion. 
Therefore, a future prototype of the system should evaluate improvements in participant's sense of presence while cards are continuously tracked using a low-latency extended Kalman filter or another method to reduce positional and rotational jitter. Another possible solution to the lack of realism due to positional and rotational tracking jitter is to periodically sample a passive haptic object's real orientation, with a long duty cycle between samples (approximately 1 second) and then smoothly transition the object to the updated orientation in a way that is imperceptible to the user. Continuous tracking would also allow the user to interact with more than one passive haptic object at a time.
%
%
In the future, our approach can be extended to include other non-card-based objects such as 3D virtual game figures, VR haptic board games, and any other VR applications that implement passive haptic objects using flat or card-based physical objects the participant can manipulate using their hands.

\section{Conclusion}

In this paper, we introduced a two-person VR experience using passive haptic feedback to give remote participants the sense of playing a game of cards together.  We showed that using physical cards can provide haptic feedback, which significantly improves standard scores of presence.  Participants reported feeling like they were playing a real game of cards with another person despite being remotely located.  We believe our system is one of the first works to present passive haptic feedback using an entire deck of cards in a VR card playing application for remotely located participants.

\section*{Acknowledgments}
The authors would like to acknowledge members of the UCSB Perceptual Engineering Lab for their feedback on early prototypes.  This work is supported in part by a PhD fellowship from Raytheon Intelligence and Space.

\bibliographystyle{abbrv-doi-hyperref}
\bibliography{CardsVR}

\end{document}


\firstsection{Tracking Algorithm}\label{supp:tracking}
\maketitle

The transformation of each marker tag from the sensor's 2D camera space to the 3D virtual world space begins with a canonical model comprised of four points (P) in model space 
\begin{math}
  (U, V, W) \in \Re ^3
\end{math} as shown in Equations \ref{eqnPt1}-\ref{eqnPt4} below.  The model coordinates represent the size and shape of each marker tag:

\begin{equation}
P_0=s\left(-\frac{1}{2}, 0, -\frac{1}{2}\right)
\label{eqnPt1}
\end{equation}

\begin{equation}
P_1=s\left(\frac{1}{2}, 0, -\frac{1}{2}\right)
\end{equation}

\begin{equation}
P_2=s\left(\frac{1}{2}, 0, \frac{1}{2}\right)
\end{equation}

\begin{equation}
P_3=s\left(-\frac{1}{2}, 0, \frac{1}{2}\right)
\label{eqnPt4}
\end{equation}

The screen space 
\begin{math}
  (x, y) \in \Re ^2
\end{math}
is related to the world space
\begin{math}
  (X, Y, Z) \in \Re ^3
\end{math}
through the camera matrix C: 

\begin{equation}
\begin{bmatrix}
x \\ y \\ 1
\end{bmatrix}
=
\begin{bmatrix}
f_x & 0 & c_x \\
0 & f_y & c_y \\
0 & 0 & 1
\end{bmatrix}
\begin{bmatrix}
X \\ Y \\ Z
\end{bmatrix}
\label{eqnCamera}
\end{equation}

The world space is then related to the model space through an affine transformation matrix, which is augmented using a Rodrigues rotation matrix obtained from a rotation vector and a translation vector:

\begin{equation}
\begin{bmatrix}
X \\ Y \\ Z
\end{bmatrix}
=
\begin{bmatrix}
R_{00} & R_{01} & R_{02} & T_x \\
R_{10} & R_{11} & R_{12} & T_y \\
R_{20} & R_{21} & R_{22} & T_z \\
\end{bmatrix}
\begin{bmatrix}
U \\ V \\ W \\ 1
\end{bmatrix}
\label{eqnObjectToWorld}
\end{equation}

Combining Equations \ref{eqnCamera} and \ref{eqnObjectToWorld}, we arrive at the Equation \ref{eqnOptimi}, which is solved for the Rotation (R) and Translation (T) terms on a per-tag basis using OpenCV's calib3d framework:

\begin{equation}
\begin{bmatrix}
x \\ y \\ 1
\end{bmatrix}
=
\begin{bmatrix}
f_x & 0 & c_x \\
0 & f_y & c_y \\
0 & 0 & 1
\end{bmatrix}
\begin{bmatrix}
R_{00} & R_{01} & R_{02} & T_x \\
R_{10} & R_{11} & R_{12} & T_y \\
R_{20} & R_{21} & R_{22} & T_z \\
\end{bmatrix}
\begin{bmatrix}
U \\ V \\ W \\ 1
\end{bmatrix}
\label{eqnOptimi}
\end{equation}

Several optimization methods were explored, with Infinitesimal Plane-based Pose Estimation (IPPE) \cite{Collins2014} producing very fast results exceeding 30 FPS. IPPE implements pose estimation over exactly 4 co-planar points, which corresponds to the four corners of the detected, square fiducial markers \cite{Collins2014}. 
We explored other methods to estimate the tag's pose in world space including Levenberg-Marquardt Optimization with iterative methods and Direct Linear Transformation (DLT) \cite{madsen2004,eade2013gauss}, but the performance was not as fast as IPPE.  Once R and T are known, objects are positioned in the corresponding world space using Equation \ref{eqnObjectToWorld}.



\section{Presence Questionnaire}\label{supp:presenceQuestionnaire}

Participants were asked to complete a set of standard presence survey, which was tailored as shown in Table \ref{tablePresence}.  Questions 6, 14, and 21 were eliminated from the standard presence questionnaire because the participant was seated and stationary throughout the study and there was no movement of the player in the VE, and question \#16 was eliminated because the participant was stationary and did not observe objects from multiple viewpoints.  The analysis results, including all subgroups, are shown in Table \ref{supp:tableMannWhitney_all} and  Figure \ref{supp:BoxPlotAll}.

\begin{table}[h]
    \centering
     \caption{Presence Questionnaire (Source: Witmer and Singer \cite{Witmer1998, Witmer2005})}
     \begin{tabular}{ccc}
         \toprule
          & Survey & Eliminated \\
         Category & Questions & Questions \\
         \midrule
         Possibility to Act & \#1, 2, 9, 10  \\
         Realism & \#3, 4, 7, 8, 18 & \#6, 14 \\
         Possibility to Examine & \#15, 24 & \#16  \\
         Self-Evaluation &  &   \\
         of Performance & \#20 & \#21  \\
         Quality of Interface & \#19, 22, 23  \\
         Passive Haptics & \#13, 17  \\
         \bottomrule
    \end{tabular}
 \label{tablePresence}
\end{table}

\section{Custom Questionnaire}\label{supp:freeFormQuestionnaire}

Participants were asked a series of free-form questions, which they completed to provide detailed feedback regarding their experiences.  These questions are listed in Table \ref{tableFreeFormQuestions}.

 \begin{table}[h]
 \centering
     \caption{Post-Survey Free-form Questionnaire Prompts}
     \begin{tabular}{cl}
         \toprule
         Question & Prompt \\
         \midrule
         Q1  & Did this virtual experience give the  \\
          & perception of playing cards with another   \\
          & person in real life? \\
         Q2 & Who do you think would most benefit from  \\
          & being able to play cards in Virtual Reality? \\
         Q3 & What did you think of playing cards in \\
          & Virtual Reality?  \\
         Q4 & What was the most appealing thing about \\
          & this experience in Virtual Reality? \\
         Q5 & What was the hardest part about using \\
          & this application to play cards? \\
         Q6 & What did you least like about the \\
          & experience? \\
         Q7 & Was there anything surprising or \\
          & unexpected about playing cards in Virtual \\
          & Reality? \\
         Q8 & Can you see yourself ever using Virtual \\
          & Reality to play cards in the future?  Why or \\
          & why not? \\
         Q9 & Why do you think someone would want to \\
          & use Virtual Reality to play cards? \\
         Q10 & What would keep people from using \\
          & Virtual Reality to play cards? \\
         \bottomrule
     \end{tabular}
     \label{tableFreeFormQuestions}
 \end{table}

\begin{table*}
\centering
\begin{minipage}[b]{.70\textwidth}
 \resizebox{1\columnwidth}{!}{%
      \begin{tabular}{cccccccc}
        \toprule
         & Possibility &  & Possibility & Self-Evaluation & Quality of & Passive \\
        Statistic & to Act & Realism & to Examine & of Performance & Interface & Haptics  \\
        
        \midrule
        P-Value & 0.172 & 0.058 & 0.111 & 0.945 & 0.739 & 0.119 \\
        Average (A) & 5.56 & 5.23 & 4.96 & 5.83 & 4.47 & 4.88 \\
        Average (B) & 5.16 & 4.72 & 5.62 & 5.90 & 4.24 & 4.29 \\
        Std Dev (A) & 0.81 & 1.02 & 1.37 & 1.14 & 1.36 & 0.83 \\
        Std Dev (B) & 1.31 & 1.41 & 1.17 & 1.04 & 1.64 & 1.28 \\
        N Samples (A) & 48 & 60 & 24 & 12 & 36 & 24 \\
        N Samples (B) & 43 & 54 & 21 & 10 & 33 & 21 \\
        Effect & 0.37 & 0.42 & 0.52 & 0.06 & 0.15 & 0.56 \\
        Power & 0.43 & 0.60 & 0.40 & 0.05 & 0.10 & 0.45 \\
        \bottomrule
        \end{tabular}
        }
        \caption{Wilcoxon-Mann-Whitney Test with PHF (A) and Without PHF (B),  including all subgroups.}
        \label{supp:tableMannWhitney_all}

\end{minipage}\qquad
\begin{minipage}[b]{.25\textwidth}
  \includegraphics[width=1\linewidth]{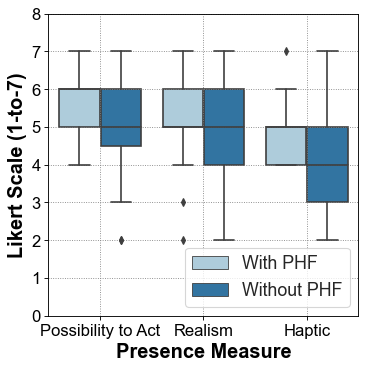}
  \captionof{figure}{Ratings for all subgroups. Interquartile range and outliers are shown.
  }
  \label{supp:BoxPlotAll}
\end{minipage}
\end{table*}

\bibliographystyle{abbrv-doi-hyperref}
\bibliography{CardsVR}